\documentclass[%
 reprint,
showpacs,preprintnumbers,
 amsmath,amssymb,
 aps,
]{revtex4-1}

\usepackage{graphicx}
\usepackage{dcolumn}
\usepackage{bm}

\DeclareMathOperator{\sech}{sech}

\begin{document}


\title{SUSY designed broken PT-symmetric optical filters}

\author{Ugo Tricoli}
\email{ugo.tricoli@onera.fr}
\author{Jean-Claude Krapez}%
\affiliation{%
 ONERA, The French Aerospace Lab, Base A\'erienne 701,13661 Salon Cedex AIR, France
}%

\begin{abstract}
		We apply the supersymmetric Darboux transformation to the optical Helmoltz wave equation to generate analytically complex-valued PT-symmetric potentials (physically a graded refractive index dielectric). PT-symmetry is then spontaneously broken controlling the amplitude of the imaginary part of the refractive index distribution. Consequently a resonance is detectable which is related to a singularity of the $S$ matrix, responsible for extraordinary high transmission and reflection peaks in the scattering spectra. We demonstrate how controlling the resonance we can achieve different amplification rates up to four order of magnitude at the exact singular point. Total transmission and very high reflection can be also obtained. All the visible portion of the spectrum can be spanned by enlarging the spatial width of the potential. All these potentials can be unified in a single device with the capability to dynamically control the imaginary part of the refractive index, thus defining a tunable dynamical optical filter behaving as a perfect amplifier, a transparent barrier or a high efficiency mirror.
		
\end{abstract}

\maketitle


\section{\label{sec:level1}Introduction}

The design of metamaterials with exotic optical properties has attracted major attention in the last decades due to the possibility of experimental realizations offered by recent technological advances \cite{cai2010optical,liu2011metamaterials}. A particularly interesting class of these materials is characterized by PT-symmetry i.e. the complex optical potential has to satisfy the necessary relation $V(r)=V^{*}(-r)$ \cite{bender1998real}. As a consequence, it is necessary for the imaginary part of the potential to be odd while the real part should be even under spatial reflection. It is well-known that Hamiltonians associated with these potentials are non-Hermitian however possessing a real spectrum because the net gain/loss distribution is zero \cite{bender1998real,miri2013supersymmetry}. In addition, PT-symmetric potentials give rise to phase transitions associated with spontaneous PT-symmetry breaking at exceptional points \cite{miri2019exceptional,ozdemir2019parity}, after which the spectrum of the eigenvalues becomes complex \cite{guo2009observation,regensburger2013observation,miri2013supersymmetry}. The eigenvalues split into complex conjugate pairs with the associated fundamental eigenfunctions been located around the maximum and minimum of the imaginary part of the optical potential related to loss and gain \cite{miri2013supersymmetry}. This behavior is typical of open systems \cite{persson2000observation} and its realization in optics is associated with many extraordinary effects \cite{miri2014susy,lin2011unidirectional,ruter2010observation,makris2011mathcal,regensburger2012parity,heinrich2014observation,longhi2018parity,el2018non}. Hence, classical optical systems with broken PT-symmetry provide the unique possibility of exploiting the special features normally associated to quantum open systems. 
In particular, a clear signature of resonant condition for the open system is when discrete real eigenvalues become complex \cite{hatano2008some}. 
Our study is directed towards the calculation of transmission and reflection spectra in order to make use of PT-symmetry to design optimal optical filters. In this framework we take into account spatial distributions of the refractive index possessing resonances that can be calculated as bound states of a Schr{\"o}dinger-like equation describing transverse modes. Hence, contrary to waveguide theory where thanks to the paraxial approximation (thus limiting the validity of the predictions for propagation directions close to the optical axis) a direct analogy is possible between optical modes and bound states of the Schr{\"o}dinger equation, here we consider scattering of waves incident along the direction of index variation (i.e. perpendicularly to the waveguide main axis, see Fig.(\ref{Fig0})). As a result, what we calculate through the Schr{\"o}dinger equation are transversal resonances of the system (and should not be considered as optically bounded due to the lack of internal total reflection). Interestingly, we found that when PT-symmetry is broken another type of resonance associated to a spectral singularity of the S-matrix is responsible for anomalous transmission i.e. the transmission coefficient becomes larger than one \cite{mostafazadeh2009spectral,mostafazadeh2015physics,shestopalov2018singularities}. This is a manifestation of the fact that the system is open towards the external environment i.e. external energy is used to generate anomalous transmission/reflection. Indeed, at the exact resonant condition (by changing the amplitude of the imaginary part of the refractive index distribution) the S-matrix has a singularity which is responsible for an amplification of both transmission and reflection of orders of magnitude.
Ultimately, the use of a broken PT-symmetric complex potential makes possible to open resonant singularities that are not obtainable with common complex potentials with only losses. Moreover, these singularities should be distinguished from another type of singularities corresponding to bound states in continuum \cite{von1993merkwurdige,marinica2008bound,regensburger2013observation,longhi2014bound,longhi2014optical,hsu2016bound}, obtainable when the waveguide is illuminated along its main axis (making possible light trapping).

Interestingly, it was proposed to use supersymmetric transformations \cite{cooper1995supersymmetry} to generate PT-symmetric potentials, however in the context of quantum mechanics making hard their experimental realization \cite{cannata1998schrodinger,sinha2002isospectral,salinas2003confluent}.
On the other hand in optics in the context of optical waveguides, it has been recently realized that supersymmetry (SUSY) can be used in order to restore PT-symmetry or to produce classes of optically equivalent potentials (in terms of reflection and transmission) with extremely different spatial distribution of optical properties \cite{miri2013supersymmetry,miri2013supersymmetric,miri2014susy}. SUSY has also been used to generate discrete real optical potentials which are reflectionless \cite{longhi2015supersymmetric} or PT-symmetric fiber optical lattices \cite{regensburger2013observation,longhi2014bound,longhi2014optical}. 
However, the spectral properties in terms of transmission/reflection spectra for perpendicular incidence on the waveguide has been investigated only with the focus on the invisibility condition   \cite{lin2011unidirectional}. Consequently, we show how a spectral singularity can be obtained via spontaneous PT-symmetry breaking of an original PT-symmetric potential generated after a single Darboux transformation of a constant potential, resulting in a simple spatial distribution of the refractive index which is possible to create in optical experiments e.g. with waveguides. As a result the final broken PT-symmetric potential can be used as an optical filter alternatively to the main use as a waveguide (i.e. for perpendicular incidence the waveguide becomes a filter). To summarize, the Darboux transformation \cite{darboux1882proposition,cannata1998schrodinger} (which constitutes the basis for supersymmetric transformations) is employed in order to generate analytical expressions of PT-symmetric potentials which are then perturbed in order to break the PT-symmetry to study the effect of phase transition on the spectral reflection and transmission of the resulting optical filters. In particular, we observe that keeping the real part of the refractive index distribution fixed (even), changing the contrast between gain and loss regions of the imaginary part of the refractive index, the same material can act as a high efficiency amplifier, reflector or a perfect transmitter in the UV and visible part of the electromagnetic spectrum.

\section{\label{sec:level1}Scattering theory and the Darboux transformation}

We consider scattering of a monochromatic electromagnetic wave by an inhomogeneous dielectric slab of finite thickness along $z$ but infinite extension in the $xy$-plane with a varying refractive index in the $z$ direction $n(z)$. The slab is surrounded by a non-absorbing background medium with refractive index $n_{b}$. Taking into account only a scalar electric field (i.e. the component $E_{x}$ corresponding to a TE-polarized wave) with the time dependence factorized as $\exp({-i\omega t})$, the curl Maxwell equations can be rewritten as the Helmholtz wave equation
\begin{equation}
\label{Helmolholtz}
\frac{\partial^2 E_{x} }{\partial z^2} + \frac{\partial^2 E_{x} }{\partial y^2} + [k_{0}n(z)]^2  E_{x} = 0,
\end{equation}
with $k_{0}=\omega/c$ the wavenumber in vacuum and $n(z)= n_b + \Delta n(z)$ the spatial dependent refractive index, with the last term being the optical contrast which is complex valued and vanishes at infinity. For a general angle of incidence the solution to Eq.(\ref{Helmolholtz}) can be factorized through an amplitude and a phase as
$
\label{sol1}
E_x(y,z) = \psi(z) \exp(ik_y y)
$,
with the incoming wavevectors related by $k_y^2= [k_{0}n_b]^2-k_z^2$. Substituting into Eq.(\ref{Helmolholtz}) and  upon addition/subtraction of the term $[k_{0}n_b]^2 \psi(z)$ we get the equation for the amplitude $\psi(z)$
\begin{equation}
\label{Schroedinger}
[-\frac{d^2 }{d z^2} + V(z)] \psi(z) = e \psi(z),
\end{equation}
which is in a Schr{\"o}dinger-like form $H\psi=e\psi$ upon definition of the optical potential as $V(z)=-k_0^2 [n(z)^2-n_b^2]$, the Hamiltonian $H=-d^2 \psi(z) /d z^2 + V(z)$ and the energy $e=k_z^2=[k_{0}n_b]^2 - k_y^2$, i.e. the square of the component of the wavevector of the incoming wave along the direction of variation of the refractive index.
%
%
Due to the reduction of the optical Helmholtz wave equation to a Schr{\"o}dinger-like equation, it is possible to apply in optics techniques originally developed in quantum mechanics (e.g. SUSY \cite{cooper1995supersymmetry,cannata1998schrodinger}). 
Indeed, SUSY transformations are based on the Darboux transform \cite{darboux1882proposition,cannata1998schrodinger} which allows building solvable profiles of the refractive index \cite{krapez2017sequences}. 
In general, the main idea of a Darboux transform is that the spectral properties of a pair of Schr{\"o}dinger Hamiltonians can be related through 
a superpotential $\sigma$ defined as the
logarithmic derivative of the solution of the Schr{\"o}dinger equation $\psi(z)$ for a potential $V(z)$ for some value of the transformation energy $\epsilon$ i.e. $\sigma(z)=\frac{1}{\psi(z)}\frac{d\psi(z)}{dz}$ 
($\epsilon$ can be complex-valued however here we restrict to real values).
The partner potential of $V(z)$ is found through the optical Darboux transform
\begin{equation}
\label{DT}
V^\prime(z) = -V(z)+ 2\sigma(z)^2 + 2\epsilon.
\end{equation}
An important remark is necessary concerning the nature of the general solution $\psi(z)$. Its analytical form must be chosen in order to avoid zeros on the real axis, thus avoiding singular superpotentials. As it was noted in \cite{cannata1998schrodinger} this can be in general obtained (as a rule of thumb in optics) taking $\epsilon<e_0$, with $e_0$ the lowest energy of the spectrum of $H$. In the quantum mechanical context this corresponds to create in the partner Hamiltonian $H^\prime$ a bound state with energy lower than all the bound states present in $H$. Consequently, the main effect of a Darboux transform is to create a partner Hamiltonian which has the same spectral properties as the original one except for the ground state which is added to the transformed Hamiltonian at the chosen energy $\epsilon$ \cite{cannata1998schrodinger}. 
Hence, a remarkable property of the Darboux transform is that it is possible to create a complex potential with bound states (that in the scattering configuration we consider here i.e. normal incidence, correspond to resonances of the refractive index distribution). Moreover, the generated potential is not necessarily PT-symmetric. In this case without symmetry, the eigenenergies are allowed to be complex valued. 
On the contrary, when PT-symmetry is preserved, the potential real/imaginary part contrast is sufficiently small thus preserving a real spectrum.
In this work we use the Darboux transform in order to generate a PT-symmetric optical potential with a single transversal resonance with real energy. 

\begin{figure}[h]
	\centering
	\includegraphics[width=8cm]{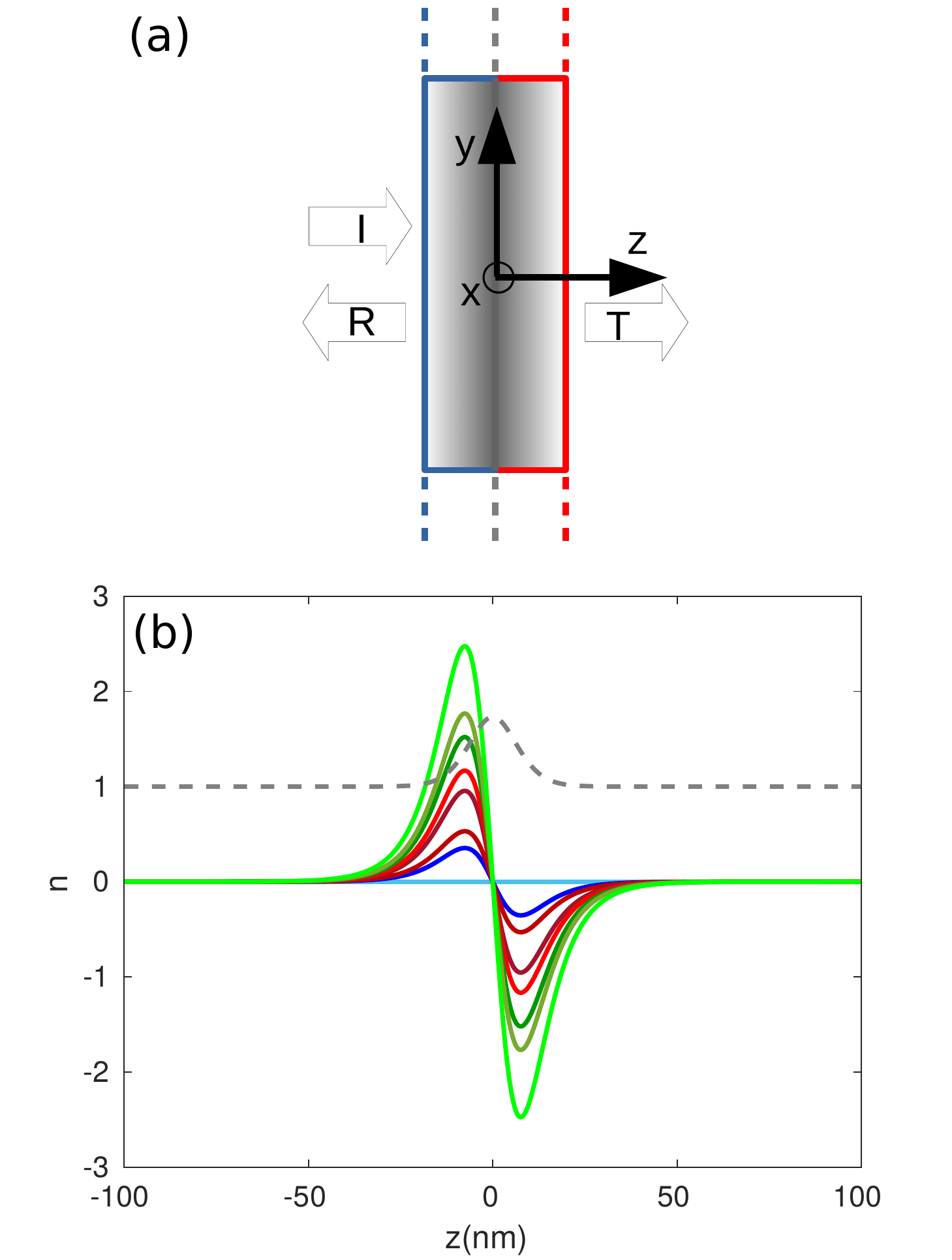}
	\caption{The refractive index distribution $n(z)$ generated after a single Darboux transform of the initial potential $V(z)=0$. (a) Scattering geometry and a physical realization of the potential $V^\prime(z)$ through a graded refractive index waveguide with gain (red contour part) and loss (blue contour part) regions. The gray area represents $\rm{Re}[n(z)]$. (b) The refractive index distribution $n(z)$ for different values of the perturbation parameter $\delta$; the color code is the same as in Fig.(\ref{Fig2}) for $\rm{Im}[n(z)]$ while the dashed gray line represents $\rm{Re}[n(z)]$.
		}
	\label{Fig0}
\end{figure}

We apply the Darboux transform to a constant initial potential that is fixed to $V=0$. We also choose $\epsilon=-0.5$ in units of $(k_s n_b)^2$ in order to create a potential with a bound state in the sense of the Schr{\"o}dinger equation (for the corresponding refractive index distributions shown in Fig.(\ref{Fig0}) we took $k_s n_b = 1/10\rm{nm^{-1}}$ with $k_s$ the scaling factor). Only combinations of hyperbolic functions (with coefficients $C_1$ and $C_2$) are used in order to avoid singularities in the superpotential. The last is a practical choice since the Schr{\"o}dinger equation with a constant potential admits also oscillating trigonometric solutions though hardly exploitable through a Darboux transform due to the presence of zeros on the real axis. All in all, a single Darboux transform can be implemented by tuning three independent parameters namely, $\epsilon$, $C_1$ and $C_2$ ($V$ is related to $\epsilon$).
Given the constraints described above, the most general solution of the Schr{\"o}dinger-like equation (\ref{Schroedinger}) with a constant potential and $e=\epsilon$ can be written as 
\begin{equation} 
\label{General_sol}
\psi(z)= C_1\exp(-\sqrt{V-\epsilon}z) + C_2\exp(\sqrt{V-\epsilon}z).
\end{equation}
We note here that reflectionless potentials as introduced in \cite{kay1956reflectionless} can be simply obtained by setting $C_1=C_2=1$ which then results into a transformed potential $V^\prime(z)\propto \sech^2(\sqrt{V-\epsilon}z)$. However, our general aim here is to generate complex valued potentials with gain/loss regions. In order to achieve this, it is necessary to choose at least one complex coefficient e.g. $C_1=1$ and $C_2=i$. Another strategy could be to choose a complex transformation energy $\epsilon$ (which is not considered here). Having fixed the general solution, the next step of a Darboux transform is the calculation of the superpotential that reads
\begin{equation} 
\label{sigma}
\sigma(z) = \sqrt{V-\epsilon} -[\sqrt{V-\epsilon}+i \sqrt{V-\epsilon}\exp(z/\sqrt{V-\epsilon})]^{-1}.
\end{equation}
Inserting it into Eq.(\ref{DT}) we get the transformed  potential 
\begin{equation} 
\label{V2}
V^{\prime}(z)  = \frac{8\lvert\epsilon\rvert}{1+\cosh(4\sqrt{\lvert\epsilon\rvert}z)} [i \sinh(2 \sqrt{\lvert\epsilon\rvert}z) -1].
\end{equation}
Interestingly, after a single Darboux transform with at least one purely imaginary coefficient we obtain a PT-symmetric potential which has an entirely real spectrum (i.e. real energy optical resonances). The associated refractive index distribution can be found through the optical potential definition i.e. $n(z)=\sqrt{n_b^2-V^\prime(z)}$, where from Eq.(\ref{V2}) the transformed potential $V^\prime(z)$ is given in units of $(k_s n_b)^2$ and $n_b=1$ for the rest of the study. In order to break in a controlled manner the PT-symmetry of the potential, we introduce an auxiliary parameter $\delta$ which is used to modify the gain/loss part of the refractive index distribution i.e. $n(z,\delta)=\rm{Re}[n(z)] + i \delta \rm{Im}[n(z)]$. The unperturbed potential $V^\prime$ as defined in Eq.(\ref{V2}) has a single real optical resonance (the associated refractive index distribution for $\delta=1$ is shown in Fig.(\ref{Fig0})). This potential can be realized in experiments by making a focused beam to impinge perpendicularly to an optical waveguide divided in two halves, one half being a gain region and the other a loss region \cite{el2007theory}. Gain can be controlled through an optical amplifier while loss can be adjusted through acoustic modulators \cite{ozdemir2019parity}. 


\section{Results: Transmission and Reflection spectra}

\begin{figure}[h]
	\centering
	\includegraphics[width=8cm]{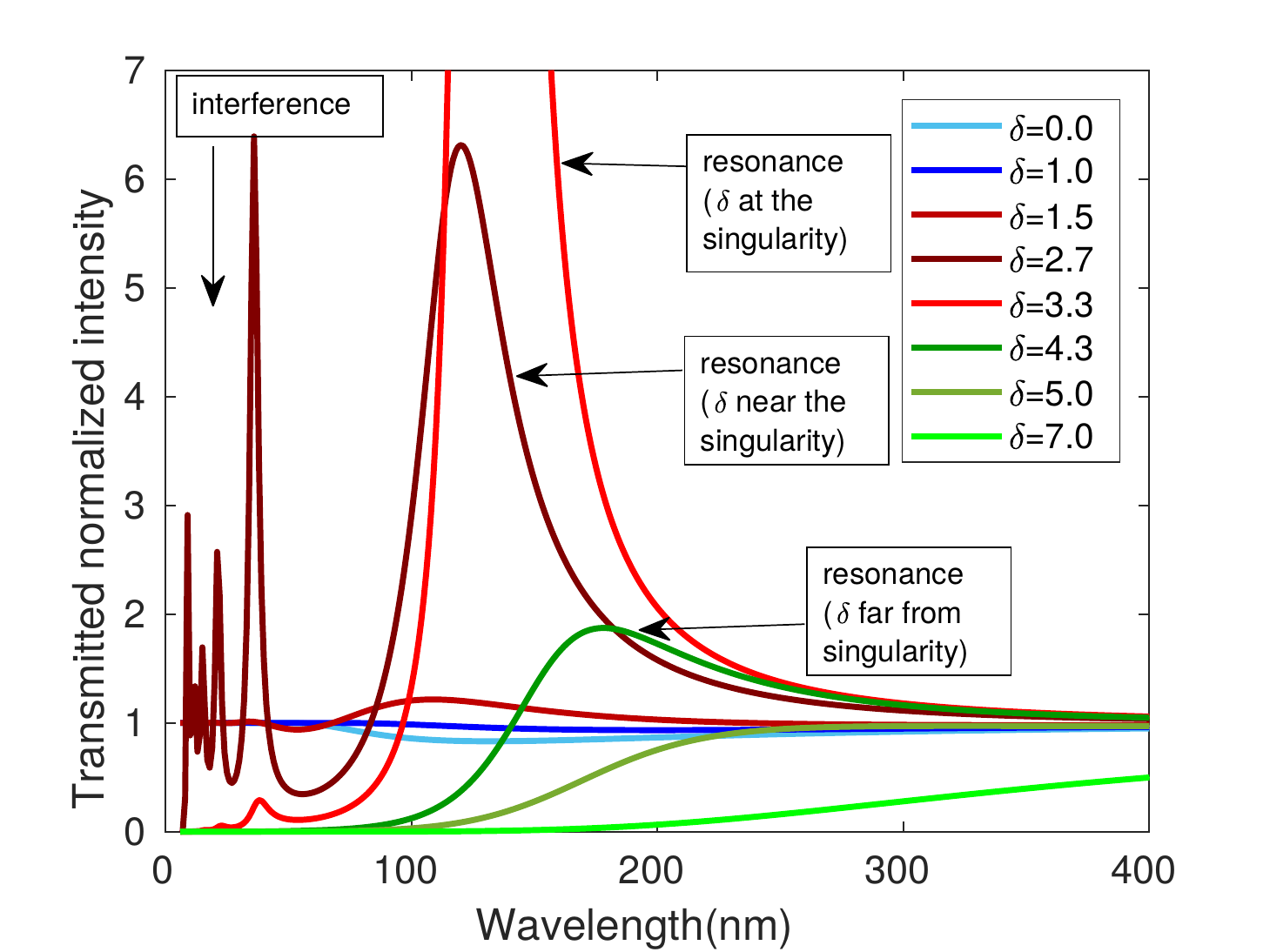}
	\caption{Transmission spectra (for incident radiation on the left side only and perpendicular incidence) of the refractive index distributions shown in Fig.(\ref{Fig0}) for different values of $\delta$. The appearance of a resonance corresponding to a spectral singularity of the S-matrix is found at $\delta=3.3$. 
	}
	\label{Fig2}
\end{figure}

\begin{figure}[h]
	\centering
	\includegraphics[width=8cm]{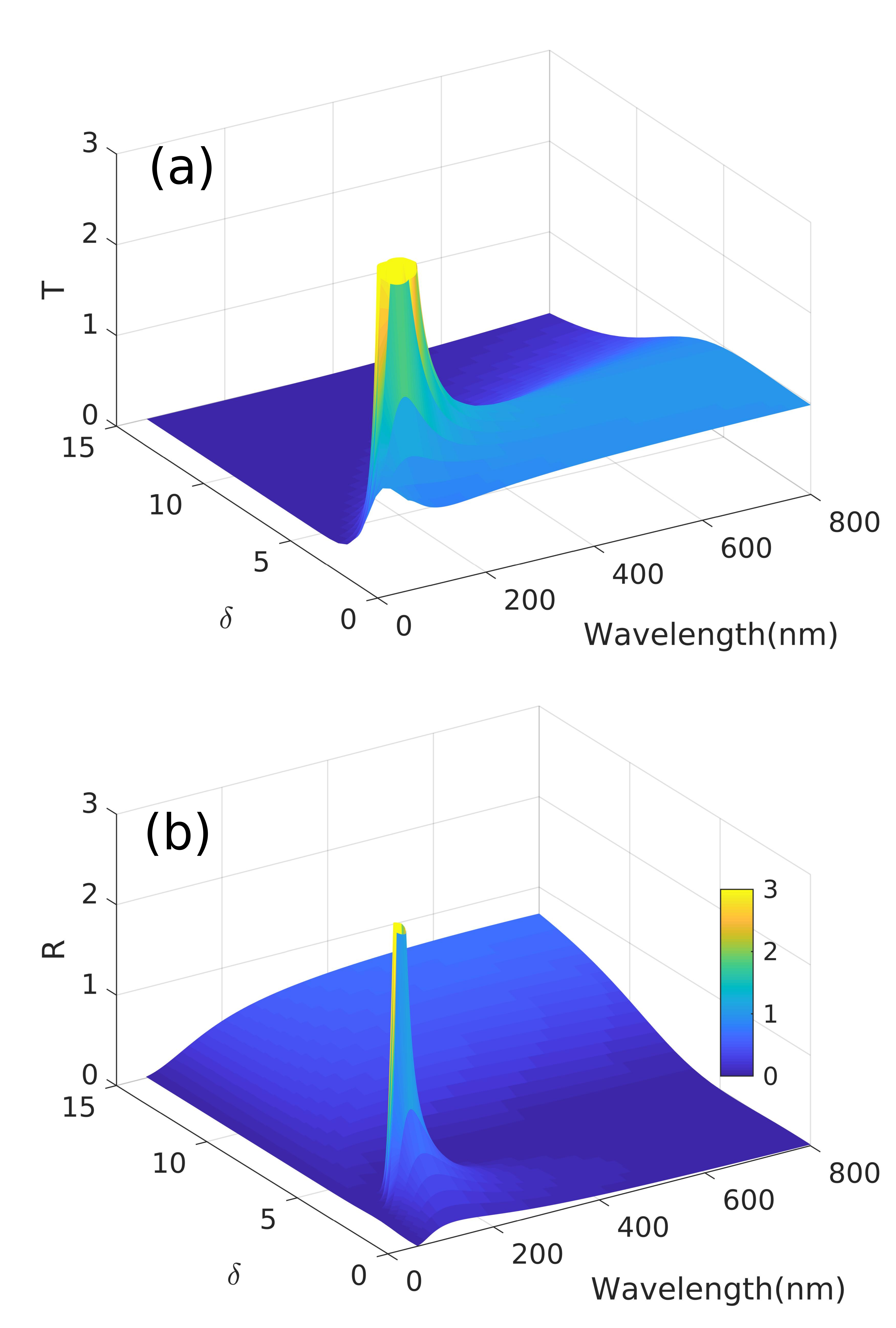}
	\caption{Transmission (a) and  reflection (b) spectra (for incident radiation on the left side only and perpendicular incidence) of various refractive index distributions obtained for different values of $\delta$ showing the appearance of a resonance corresponding to a spectral singular point of the S-matrix for $\delta=3.3$. 
	}
	\label{Fig3}
\end{figure}

Using a standard numerical implementation of the optical transfer matrix method \cite{lekner1994light,yeh1977electromagnetic,pascoe2001reflectivity} we calculate in Fig.(\ref{Fig2}) the transmission spectra associated to the refractive index distribution $n(z,\delta)$ (for perpendicular incidence on the left side only).
As noted in \cite{ge2012conservation}, a 1D PT-symmetric potential can be viewed from the two sides differently, one side being a gain side while the other one a loss side 
and consequently a generalized energy conservation relation was therein derived.
Moreover, 
the presence of a resonance (corresponding to a real bound state of the Schr{\"o}dinger equation) can be detected as a unitary transmission peak at the corresponding eigenenergy \cite{neufeld2018calculating}. 
In accordance, in Fig(\ref{Fig2}) the PT-symmetric potential with $\delta=1$ exhibits the maximal transmission (the peak expected in the energy space becomes very wide in direct $\lambda$ space) at the expected location $\lambda= 2 \pi / k_s \sqrt{1/ \lvert e \rvert} \approx89\rm{nm}$ (for $e=-0.5$ in units of $k_s^2$).
Then, for longer wavelengths transmission falls off to values lower than one. In the same spectral region (see Fig.(\ref{Fig3})), reflection is increased but the net energy balance (on the left side only) is positive indicating a gain effect (correspondingly the absorption coefficient becomes negative). This is an evidence of the opening of the system toward the external environment (as it is known for quantum open systems \cite{persson2000observation,hatano2008some,garmon2015bound}). The dielectric system is using external energy in order to preserve the resonance at the prescribed energy. Then, for longer wavelengths the potential is becoming progressively transparent due to its limited spatial extension (see Fig.(\ref{Fig3})). 
%
We then explore how the transmission spectra can be altered by perturbing through $\delta$ the PT-symmetric potential $V^\prime$ created after a Darboux transform.
Interestingly, at $\delta=1.24$ i.e. just before the PT symmetry breaking point, the system has two resonances at approximately the same real energy thus preserving the PT symmetry.
Spontaneous PT-symmetry breaking is reached at $\delta= 1.245$ (for the chosen  Darboux transform energy $\epsilon=-0.5$). Above this exceptional point the spectrum of the resonances of the potential abruptly becomes complex. As noted before, the transversal resonance splits into a complex conjugate pair of resonant states peaked around the maximum and minimum of the imaginary part of the potential \cite{miri2013supersymmetry}. 
%
Remarkably, as soon as PT-symmetry is broken transmission becomes anomalous (see the cases with $\delta=1.5$ and $\delta=2.7$ in Fig.(\ref{Fig2})). 
Increasing further the amplitude of the imaginary part of the potential, at $\delta=3.3$ we find an anomalous increase of four order of magnitude of the transmission coefficient.
Moreover, the potential is behaving as an optical amplifier for both transmitted and reflected radiation in the far UV around the resonance at $\lambda= 131 \rm{nm}$ (see Fig.(\ref{Fig3})).
This behavior can be explained as the manifestation of a resonance connected to a spectral singularity of the optical S-matrix.
However, the spectral position of the corresponding transmission/reflection resonant peaks is not corresponding to the real part of an eigenvalue, but it can be estimated by a study of singular points of the $S$ matrix \cite{mostafazadeh2009spectral,hatano2008some} (see the Appendix).
Differently, the narrow peaks for very short wavelengths which are amplified due to the complex nature of the potential are linked to scattering interferences of geometrical origin caused by the real part of the potential acting as an optical cavity. 
Increasing further more $\delta$, the transmission is progressively reduced and only longer wavelengths maintain an almost reflectionless total transmission. In fact, far (in terms of $\delta$) from the spectral singular point, the resonant peak is shifted to larger wavelengths and becomes so broad that the amplifying effect is very limited. However, the barrier is reflectionless and the transmission goes to unity in all the visible above $\lambda \approx 300 \rm{nm}$ thus giving a transparent potential. For very large $\delta>10$ the potential becomes almost completely reflective with no transmission in the visible spectrum. 
Consequently, the spectrum of a very large purely real barrier is recovered i.e. transmission is suppressed and reflection is high in all the visible. We conclude by noting that the potential seen from the left side can be used as a very efficient optical amplifier and all the visible portion of the spectrum can be spanned by expanding the spatial width of the potential (by changing the scaling factor) thus changing the energy of the spectral singular point and consequently the spectral position of the resonance in the broken PT-symmetry regime (see the Appendix).

Finally, all these results can be unified if we think of a device as schematically depicted in Fig.(\ref{Fig0}) through which the distribution of gain and loss can be controlled dynamically. Having this, it is possible to produce a material which acts as an optical barrier with a fixed real part while, the imaginary part can be dynamically changed to achieve different spectral functionalities. In this way the same material could be used as an amplifier, a transparent filter or a high efficiency mirror in the UV and visible portion of the spectrum.

\bibliography{DTfilters}

\newpage
\appendix

\section{Correspondence between bound states and resonances}

The Darboux transform has been used to generate potentials for the Schr{\"o}dinger equation. In particular, one single Darboux transform is used to create a potential which supports a single bound state in the sense of negative eigenvalue of the Schr{\"o}dinger equation. In optics, in the context of the Helmholtz equation, the potential is related to the spatial variation of the refractive index i.e. $V=-k_0^2(n(z)^2-n_b^2)$. Importantly, a negative potential corresponds to a positive refractive index variation or equivalently bound states of the Schr{\"o}dinger equation corresponds to optical resonances of the Helmoholtz wave equation. In the context of waveguides, it has been possible to achieve a physical realization of bound states in optics (in the sense of the Schr{\"o}dinger equation i.e. a wave at the eigenenergy is trapped in the potential without any possibility to escape) by exploiting interference and notably the total internal reflection for some incident angles. This means that optically the state (or the mode) is bounded only if the incident angle is larger than the critical angle. Hence, only in the limit of fields propagating close to the waveguide main axis it is possible to speak about light being bounded. Indeed, the proper description of optical bound states passes through the paraxial approximation to the general Helmholtz wave equation and is valid only close to the optical axis. On  the other hand, if light is sent perpendicularly to a waveguide, transversal resonances are excited and can be identified by transmission measurements i.e. total transmission is expected at the exact location of the resonances \cite{neufeld2018calculating} (on the contrary, for illumination along the waveguide main axis, transversal resonances turn to be optical modes propagating along the waveguide). In the case of broken PT-symmetric potentials the net effect of the potential (from one side only) on the field can lead to gain effects that result in transmission larger than one. 
Remarkably, a broken PT-symmetric potential can open spectral singularities of the S-matrix that cause large amplification of the incoming radiation. Hence, the broken PT-symmetric waveguide illuminated perpendicularly acts as an active cavity.


\section{Transmission and Reflection spectra}

\begin{figure}
	\centering
	\includegraphics[width=8cm]{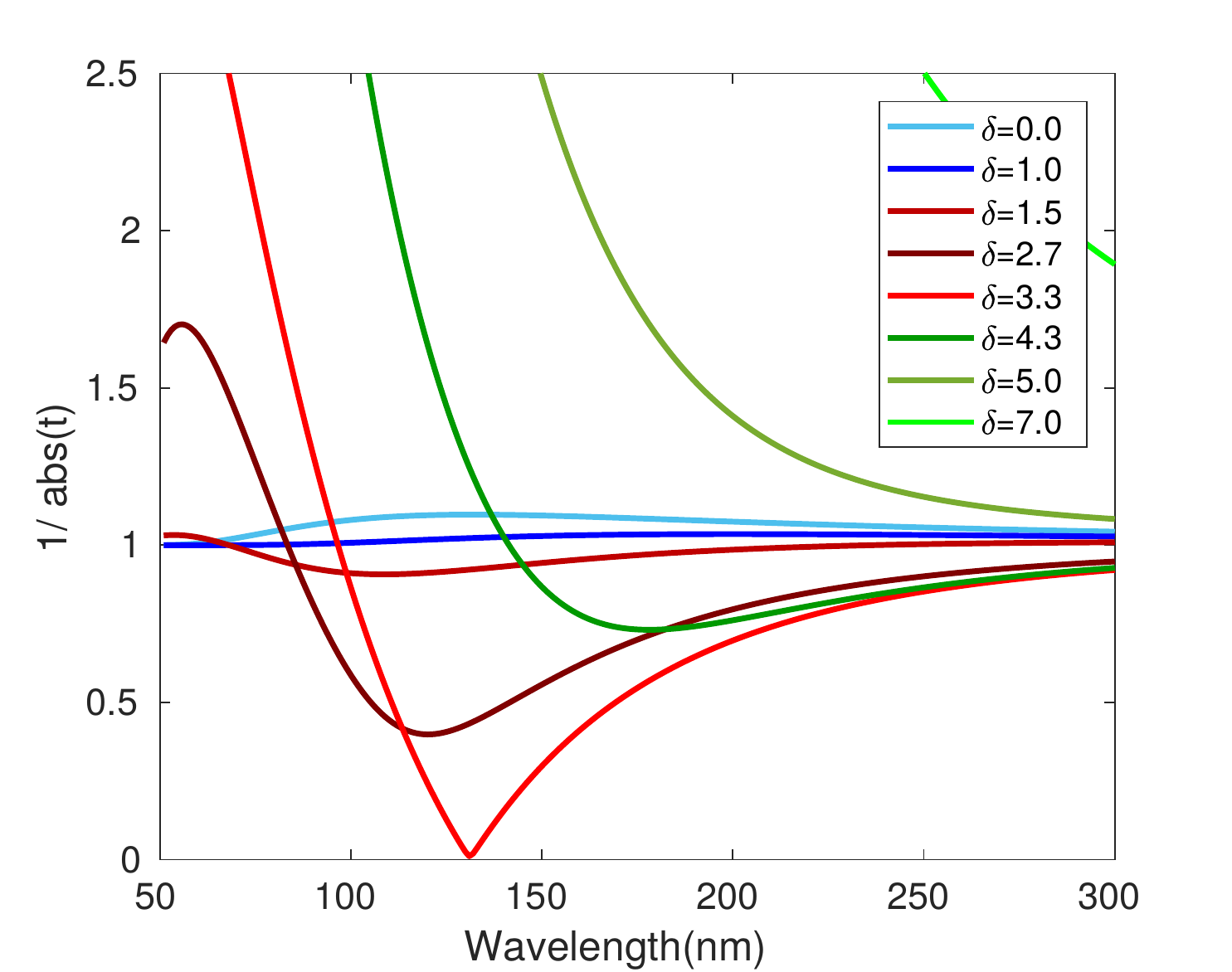}
	\caption{The inverse of the module of the transmission amplitude of various refractive index distributions obtained for different $\delta$s. A zero is identified at $\lambda=131\rm{nm}$ for $\delta=3.3$. The incident radiation is on the left side only and with perpendicular incidence.
	}
	\label{Fig2c}
\end{figure}

In Fig.(\ref{Fig2c}) we study the inverse of the module of the transmission amplitude in order to find and locate the spectral singular points of the optical $S$ matrix. These singular points corresponds to zeros of the denominator of the transmission and reflection amplitudes which cause in principle infinitely large transmission and reflection coefficients. For the definition of the transmission and reflection amplitudes (and coefficients) and of the transfer matrix we refer to \cite{pascoe2001reflectivity}.

\begin{figure*}[h]
	\centering
	\includegraphics[width=7.5cm]{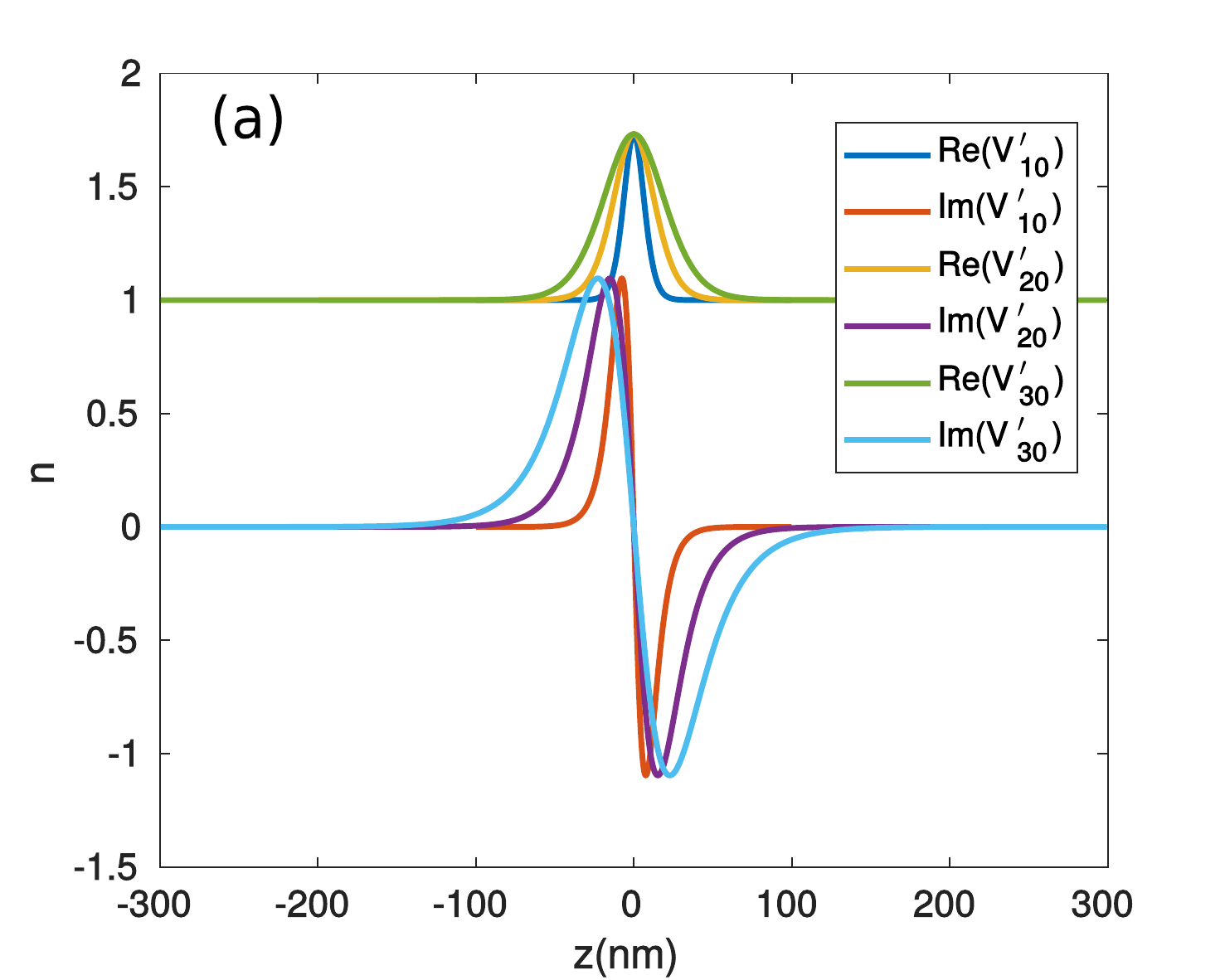}
	\includegraphics[width=7.5cm]{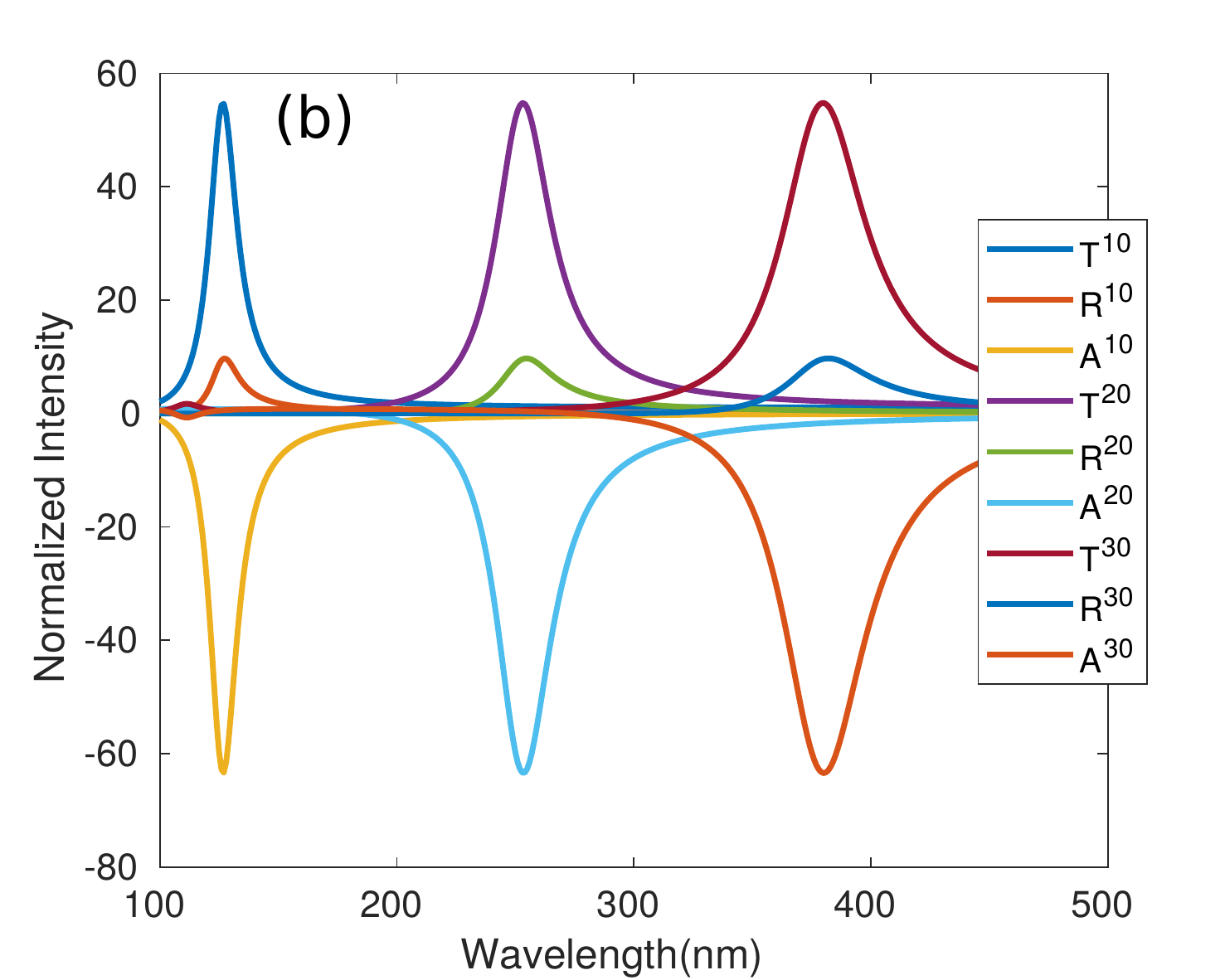}
	\caption{(a) The refractive index distribution $n(z)$ obtained with $\delta=3.1$ and enlarged  along the direction $z$ through the scaling factor $k_s  =1/10,1/20,1/30\rm{nm^{-1}}$. (b) The left transmission, reflection and absorption spectra (for perpendicular incidence) corresponding to an amplifier from the UV to VIS obtained for the potentials shown in (a).
	}
	\label{Fig2b}
\end{figure*}

The refractive index distribution associated to the potential $V^\prime(z)$ seen from the left side can be used as a very efficient optical amplifier and all the visible portion of the spectrum can be spanned by expanding the spatial width of the potential thus changing the spectral position of the singularity linked to the resonance in the broken PT-symmetry regime (see Fig.(\ref{Fig2b})).

\begin{figure*}[h]
	\centering
	\includegraphics[width=7.5cm]{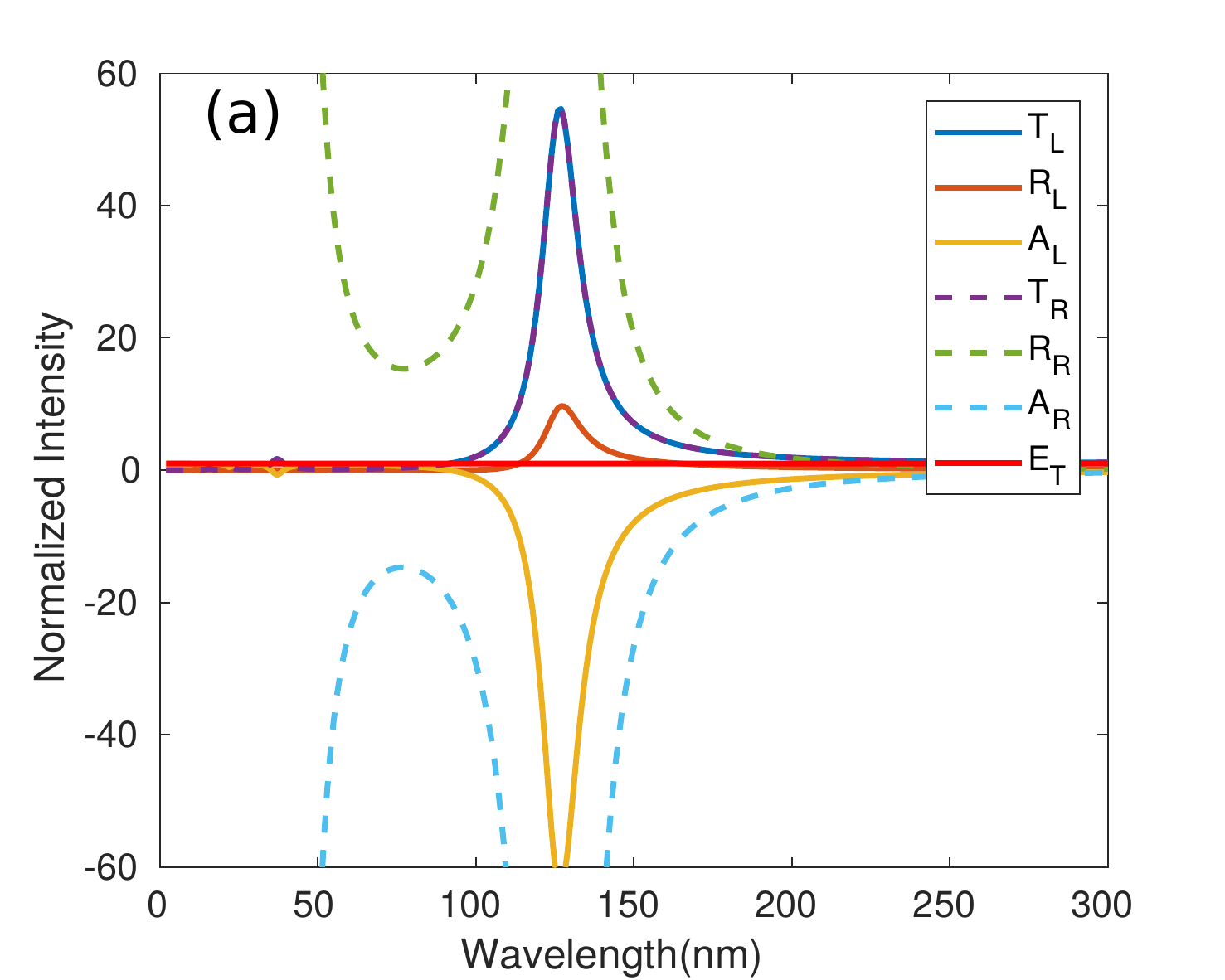}
	\includegraphics[width=7.5cm]{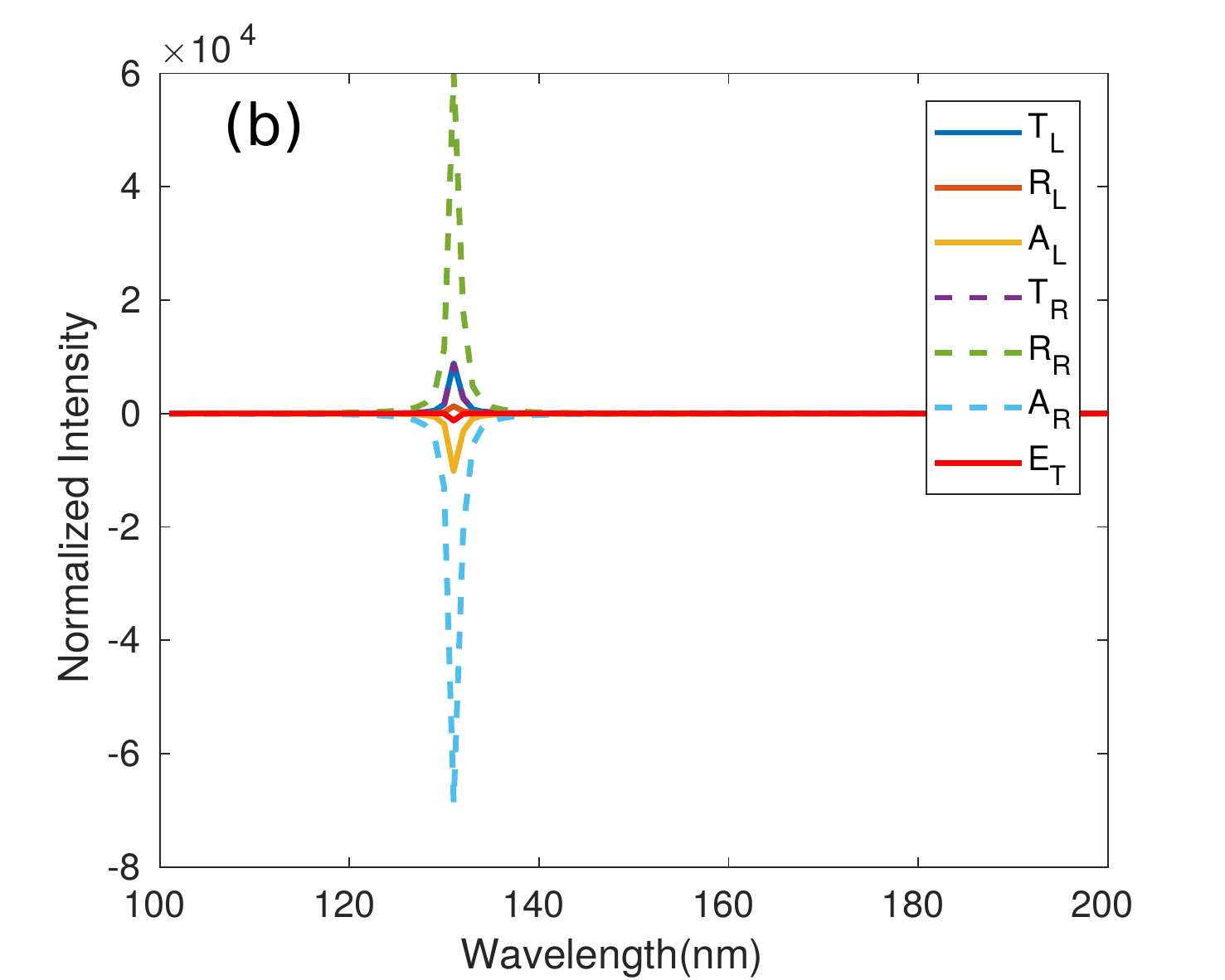}\\
	\includegraphics[width=7.5cm]{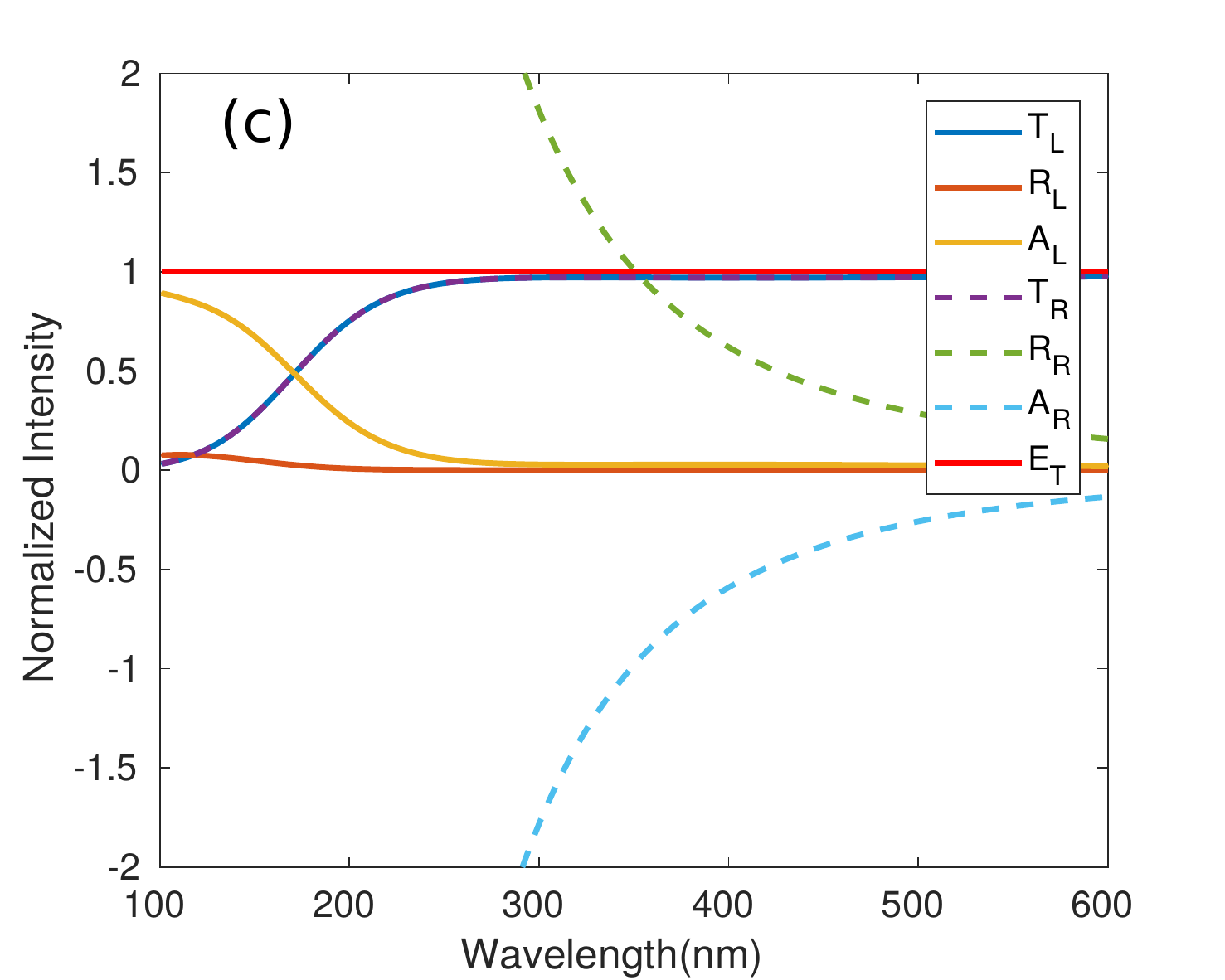}
	\includegraphics[width=7.5cm]{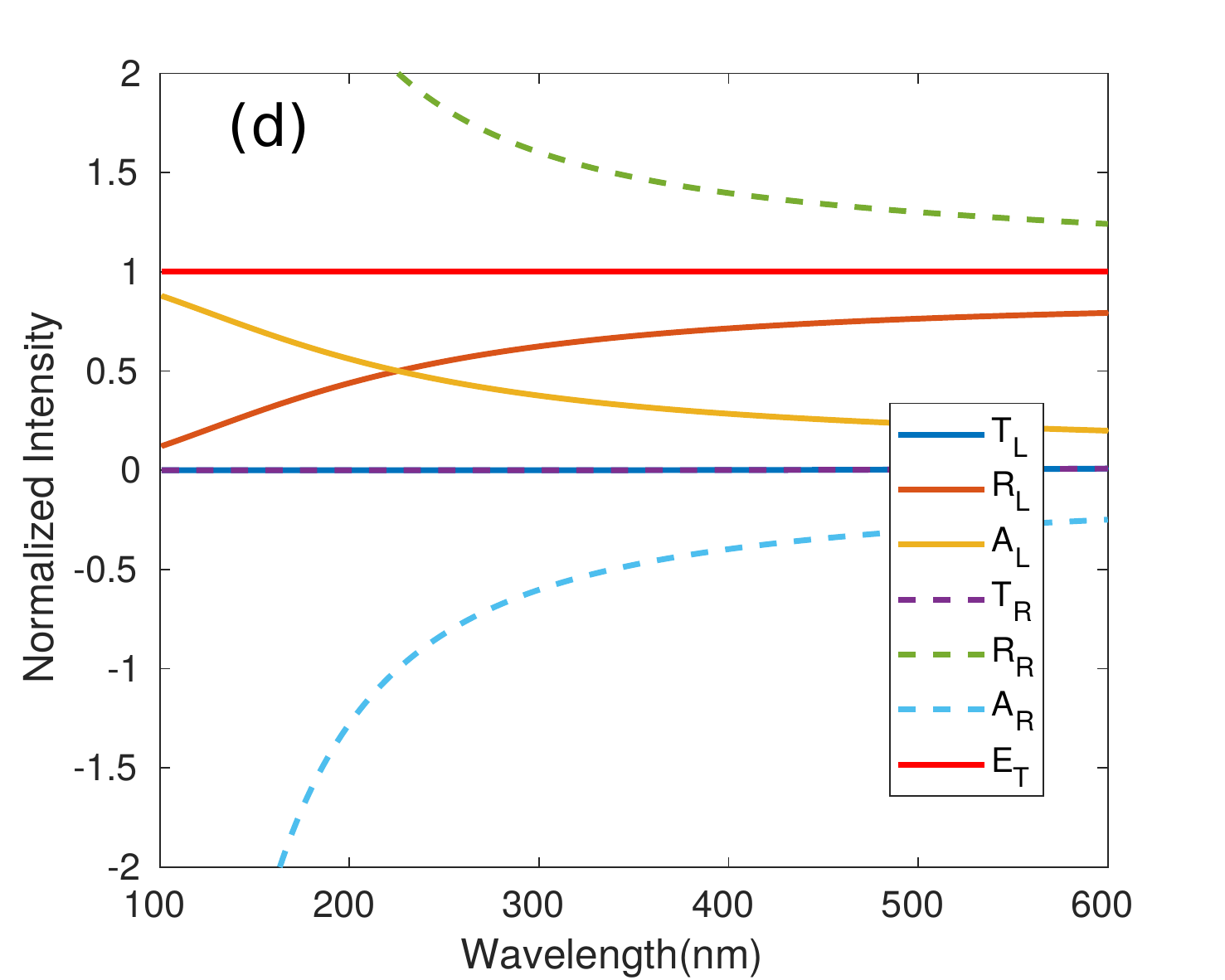}
	\caption{The left (L) and right (R) transmission, reflection and absorption spectra (for perpendicular incidence) associated to the refractive index distribution $n(z)$ generated after a single Darboux transform of $V=0$ and its perturbations causing a PT-symmetry breaking through different $\delta\rm{s}$. (a) An amplifier in the UV obtained with $\delta=3.1$ associated with a resonance for $\delta$ close to a spectral singularity. (b) A perfect amplifier in the UV obtained with $\delta=3.3$ associated with a resonance for $\delta$ at a spectral singularity. (c) A transparent barrier in the visible obtained with $\delta=5$. (d) A mirror in the visible obtained with $\delta=20$. The red line corresponds to the total energy which is conserved $E_{T}=R_{L}R_{R}+2T-T^2$ \cite{ge2012conservation}.	
	}
	\label{Fig4}
\end{figure*}

In the main text we have considered mainly transmission spectra from the left side only to elucidate how PT-symmetry breaking can be used to design exceptional optical filters. In Fig.(\ref{Fig4}) we study how just changing the imaginary part of the refractive index distribution we can obtain filters with very different properties considering also reflection from both left and right sides. In practice Fig.(\ref{Fig4}) explains details of Fig.(3) in the main text considering the spectra for illumination on both sides.
Surprisingly, the potential is behaving as an optical amplifier for both transmitted and reflected radiation in the UV around the resonance at $\lambda\approx 131 \rm{nm}$ for $\delta=3.1$. The same is true at the singularity $\delta=3.3$ where we note also that reflection from the right side is amplified maximally. Further, increasing $\delta$ to $5$ see Fig.(\ref{Fig4}c), the complex barrier becomes completely transparent in the visible portion of the spectrum. Differently from the previous cases, much of the radiation in the UV is absorbed. 
However, the barrier is reflectionless and the transmission goes to unity in all the visible above $\lambda \approx 300 \rm{nm}$ thus giving a transparent potential.
For extremely large $\delta=20$ (see Fig.(\ref{Fig4}d)), 
the spectrum of a very large purely real barrier is recovered i.e. transmission is suppressed and almost unit reflection is recovered for all the visible.

\end{document}